\documentclass[prd,aps,twocolumn,amsmath,amssymb,nofootinbib,preprintnumbers]
{revtex4}

\usepackage{graphicx}
\usepackage{dcolumn}
\usepackage{bm}
\usepackage{amsmath}
\usepackage{amsfonts}
\usepackage{euscript,bbm}
\usepackage{ifthen}
\usepackage{psfrag}
\usepackage{slashed}

\def\ls{\mathrel{\lower4pt\vbox{\lineskip=0pt\baselineskip=0pt
           \hbox{$<$}\hbox{$\sim$}}}}
\def\gs{\mathrel{\lower4pt\vbox{\lineskip=0pt\baselineskip=0pt
           \hbox{$>$}\hbox{$\sim$}}}}
\def\drawbox#1#2{\hrule height#2pt

\hbox{\vrule width#2pt height#1pt \kern#1pt
              \vrule width#2pt}
              \hrule height#2pt}

\def\Asym#1#2{\vcenter{\vbox{\drawbox{#1}{#2}
              \kern-#2pt       
              \drawbox{#1}{#2}}}}

\newcommand{\be}{\begin{equation}}
\newcommand{\ee}{\end{equation}}
\newcommand{\bea}{\begin{eqnarray}}
\newcommand{\eea}{\end{eqnarray}}

\newcommand{\gsim}{\lower.7ex\hbox{$\;\stackrel{\textstyle>}{\sim}\;$}}
\newcommand{\lsim}{\lower.7ex\hbox{$\;\stackrel{\textstyle<}{\sim}\;$}}

\begin{document}

%
\title{Successful Supersymmetric Dark Matter with Thermal Over/Under-Abundance from Late Decay of a Visible Sector Scalar}

\author{Rouzbeh Allahverdi$^{1}$}
\author{Bhaskar Dutta$^{2}$}
\author{Kuver Sinha$^{2}$}

\affiliation{$^{1}$~Department of Physics and Astronomy, University of New Mexico, Albuquerque, NM 87131, USA \\
$^{2}$~Mitchell Institute of Fundamental Physics and Astronomy, Department of Physics and Astronomy, Texas A\&M University, College Station, TX 77843-4242, USA}

\begin{abstract}
We present an explicit model where the decay of an $R$-parity even scalar $S$ with ${\cal O}({\rm TeV})$ mass is the origin of non-thermal dark matter. The correct relic abundance can be produced for both large and small annihilation rates in accordance with the Fermi constraints on the annihilation cross-section. This scenario has advantages over that of non-thermal dark matter from modulus decay. First, branching ratio for production of $R$-parity odd particles can be made quite small by a combination of $S$ couplings to matter fields and kinematic suppression, enabling us to obtain the observed dark matter relic density in cases of thermal underproduction as well as overproduction. Second, gravitino production is naturally suppressed by the virtue of decaying scalar belonging to the visible sector. The decaying scalar can also successfully generate baryon asymmetry of the universe, and may provide an explanation for the baryon-dark matter coincidence puzzle.
\end{abstract}

MIFPA 12-46\\ December, 2012
\maketitle

\section{Introduction}

Weakly interacting massive particles (WIMPs) are promising dark matter (DM) candidates that can explain the DM relic abundance, as precisely measured by cosmic microwave background experiments~\cite{WMAP}, via thermal freeze-out of annihilation in the early universe. The nominal DM annihilation rate for this scenario, called the WIMP miracle, is $\langle \sigma_{\rm ann} v \rangle \approx 3 \times 10^{-26}$ cm$^3$ s$^{-1}$.

WIMPs typically arise in models of particle physics beyond the standard model (SM). In supersymmetric (SUSY) models with conserved $R$-parity, the lightest supersymmetric particle (LSP) is the DM candidate. The lightest neutralino is the most suitable candidate with prospects for detection in various direct and indirect searches. However, in large regions of SUSY parameter space the thermal relic abundance of the LSP is different from the observed value. Typically, if the lightest neutralino is a Higgsino or Wino, the annihilation cross-section is large compared to the nominal value, while for a Bino the annihilation cross-section is  small.

A mainly Higgsino LSP is motivated by considerations of naturalness ~\cite{Papucci:2011wy, Hall:2011aa, Baer:2012uy, Allahverdi:2012wb, Gogoladze:2012yf, Baer:2012cf}. If the Higgsino mass is in the sub-TeV region, the annihilation rate is typically larger than $3 \times 10^{-26}$ cm$^3$ s$^{-1}$, thus resulting in an insufficient thermal relic abundance. The Fermi-LAT data \cite{GeringerSameth:2011iw} constrains the DM annihilation rate at the present time. These bounds do not allow much room for too large an annihilation cross-section. Further, for lower values of the LSP mass, Fermi-LAT data appears to prefer smaller annihilation cross-sections. This prefers Bino type LSP, which will have a thermally overproduced abundance.

Therefore, since the presently motivated DM annihilation cross-sections are either larger or smaller than the nominal thermal freeze-out value, one is naturally led to consider non-thermal scenarios to obtain the correct DM relic density. A late-decaying field $S$ that reheats the universe below the freeze-out temperature $T_{\rm f} \sim m_\chi/25$ ($\chi$ denoting the DM particle) can provide a non-thermal origin for DM. If the annihilation rate is larger than the nominal value, the LSPs produced from $S$ decay undergo residual annihilation before reaching their final abundance. Models that prefer Higgsino DM can be accommodated in such a scenario. If the number density of LSPs produced from $S$ decay is very small and/or the annihilation rate is smaller than the nominal value, annihilation will be inefficient and the final DM relic density will the same as that from $S$ decay. This scenario can accommodate both types of models with Higgsino and Bino DM.

A gravitationally coupled modulus field, typically arising in SUSY and superstring-inspired models \cite{BBN1}, serves as a standard candidate for the field $S$. Moduli that are heavier than 50 TeV decay before the onset of big bang nucleosynthesis (BBN), and hence will not ruin its successful predictions on the primordial abundance of light elements. The decay also provides a non-thermal origin for DM production \cite{Moroi:1999zb, Dutta:2009uf}.
However, in such a scenario, the branching ratio for gravitino production from late decay (denoted by ${\rm Br}_{3/2}$) must be sufficiently suppressed. Otherwise, the decay of gravitinos will lead to DM overproduction or, for gravitino mass below $40$ TeV, will ruin the success of BBN. It is not easy to satisfy this requirement in scenarios where modulus decay is the source of DM production. For example, in simple models such as KKLT \cite{Kachru:2003aw} ${\rm Br}_{3/2}$ turns out to be too large by a factor of $\sim 10^3$ \cite{Allahverdi:2010rh}.
Furthermore, it is difficult to obtain a sufficiently small DM abundance from moduli decay without residual annihilation. Even if one suppresses LSP production from two-body decays of moduli, the three-body decays will still be unacceptably large~\cite{Allahverdi:2010rh}.
Suppressing the total decay rate of the modulus can help the situation, but it involves non-trivial conditions on the K\"ahler geometry of the underlying effective supergravity theory.

In this paper we explore the late decay of a visible sector scalar field $S$ that serves as the origin of non-thermal DM. It has the following advantages: $(1)$ Gravitino production from $S$ is naturally suppressed for gravitino mass as low as ${\cal O}({\rm TeV})$ by virtue of the fact that $S$ is a visible sector field and $(2)$ Due to kinematic suppression, $R$-parity odd particles are produced at one-loop or two-loop level, which can yield ${\rm Br}_{\chi} \ll 10^{-3}$. This immediately opens up the option of obtaining non-thermal DM purely from branchings of the scalar field, without undergoing further annihilation. As we discussed, for Bino LSP, this is the only option. For Higgsinos, this allows both options of a relic density set by pure branching ratios, as well as by annihilation.

We present an explicit model where a visible sector scalar $S$ with a mass $m_S \sim {\cal O}({\rm TeV})$ and $R$-parity charge $+1$ has direct couplings to new colored fields $X,~{\bar X}$. If $S$ is lighter than $X,~{\bar X}$ particles, as well as all colored superparticles, it dominantly decays into two gluons at one-loop level. $S$ decay also produces DM particles via loops, but this mode is suppressed by powers of $m_\chi/m_S$ and $g_{1,2}/g_3$, with $g_{1,2,3}$ being the $U(1)_Y,~SU(2)_W,~SU(3)_C$ gauge couplings respectively. As we will see, this model can yield the correct DM relic abundance for of the thermal underproduction and overproduction cases, and also lead to successful late-time baryogenesis.

The paper is organized as follows. In Section II, we briefly review non-thermal DM form modulus decay and the associated problems. In Section III, we introduce a model with late decaying scalar field and discuss its properties. In Section IV, we show that this model can successfully explain the DM content for both large and small annihilation cross sections. In Section V, we discuss baryogenesis in this model and point out that the presented non-thermal DM scenario can also address the baryon-DM coincidence puzzle. We conclude the paper in Section VI.

\section{Non-thermal Dark Matter from a Late Decaying Scalar Field}

In this section, we discuss the various possibilities for non-thermal production of DM from a late decay.

\subsection{Reheating by late decay}

We consider a scalar field $S$ with mass $m_S$ and decay width $\Gamma_S$. Assuming that $S$ has acquired a large vacuum expectation value during inflation, it will start oscillating about the minimum of its potential with an initial amplitude $S_0$ when the Hubble expansion rate is $H \sim m_S$. Oscillations of $S$ behave like matter, with an initial energy density $\rho_S = m^2_S S^2_0/2$. The energy density of the universe at this time, dominated by thermal bath, is $\rho_{\rm r} = 3 m^2_S M^2_{\rm P}$.

The quantity $\rho_S/\rho_{\rm r}$ is redshifted $\propto a$, with $a$ being the scale factor of the universe. After using the fact that $H$ is redshifted $\propto a^{-2}$ for a radiation-dominated universe, we find the necessary condition for $S$ to be dominant at the time of decay
\be \label{domination}
{S_0 \over M_{\rm P}} \gg \left({\Gamma_S \over m_S}\right)^{1/4} .
\ee
Decay of $S$ reheats the universe to a temperature $T_{\rm r} \sim (\Gamma_S M_{\rm P})^{1/2}$. As a numerical example, for $m_S \sim {\cal O}({\rm TeV})$ and $T_{\rm r} \sim 3$ MeV (in order to be compatible with BBN), Eq.~(\ref{domination}) implies $S$ dominance for $S_0 \gg 10^{13}$ GeV.

If $S$ dominates the universe at a temperature $T_{\rm dom}$, we will have
\bea \label{entropy}
{\rho_{\rm r, after} \over \rho_{\rm r, before}} & = & {T_{\rm dom} \over T_{\rm r}} \, , \nonumber \\
&& \, \nonumber \\
{s_{\rm after} \over s_{\rm before}} & = & \left({T_{\rm dom} \over T_{\rm r}}\right)^{3/4} ,
\eea
where ``before'' and ``after'' are in reference to the epoch of $S$ decay, and we have used the fact that $\rho_{\rm r, after} = \rho_S$.

It is seen from Eq.~(\ref{entropy}) that $S$ decay releases a large entropy that dilutes any pre-existing quantity in the thermal bath. For the above numerical example where $m_S \sim {\cal O}({\rm TeV})$ and $T_{\rm r} \sim 3$ MeV, the entropy release factor can be as large as $10^8$.

\subsection{Dark matter from late decay}

Provided that $T_{\rm r} < T_{\rm f} \sim m_\chi/25$, decay of $S$ will dilute any thermally produced DM by a large factor as mentioned above. However, $S$ decay itself produces DM particles. The abundance of non-thermally produced DM is given by
\be \label{nonthr}
{n_\chi \over s} = {\rm min} ~ \left[Y_S ~ {\rm Br}_\chi ~ , ~ \left({n_\chi \over s}\right)_{\rm thr} ~ \left({T_{\rm f} \over T_{\rm r}}\right) \right] .
\ee
Here $Y_S \equiv 3 T_{\rm r}/4 m_S$, ${\rm Br}_\chi$ is the branching fraction for production $R$-parity odd particles from $S$ decay, and $(n_\chi/s)_{\rm thr})$ denotes DM abundance obtained via thermal freeze-out that is related to the observed DM relic abundance $(n_\chi/s)_{\rm obs}$ through:
\bea \label{thr}
\left({n_\chi \over s}\right)_{\rm thr} & = & \left({n_\chi \over s}\right)_{\rm obs} ~ {3 \times 10^{-26} ~ {\rm cm}^3 ~ {\rm s}^{-1} \over \langle \sigma_{\rm ann} v \rangle} \, , \\
& & \, \\ \nonumber
\left({n_\chi \over s}\right)_{\rm obs} & \approx & 5 \times 10^{-10} ~ \left({1 ~ {\rm GeV} \over m_\chi}\right) \,  .
\eea

The abundance of DM particles immediately after their production from $S$ decay is given by $Y_S {\rm Br}_\chi$. If $n_\chi \langle \sigma_{\rm ann} v \rangle < H(T_{\rm r})$, DM annihilation will be inefficient at temperature $T_{\rm r}$. In this case, the final DM relic abundance will be given by the first inside the brackets in Eq.~(\ref{nonthr}). On the other hand, if $n_\chi \langle \sigma_{\rm ann} v \rangle > H(T_{\rm r})$, annihilation will be efficient right after $S$ decay. This will somewhat reduce the abundance of DM particles produced from $S$ decay, in which case the final relic density will be given by the second term inside the brackets in Eq.~(\ref{nonthr}).

There are therefore two possible scenarios for obtaining the correct DM relic density from $S$ decay:
\begin{itemize}
\item{
{\bf Annihilation Scenario}: If $\langle \sigma_{\rm ann}\rangle > 3 \times 10^{-26} ~ {\rm cm}^3 ~ {\rm s}^{-1}$, then $(n_\chi/s)_{\rm thr} < (n_chi/s)_{\rm obs}$ (hence ``thermal underproduction''). The large annihilation cross section can reduce the abundance of DM particles produced from $S$ decay to an acceptable level, provided that:
\be \label{anncond}
T_{\rm r} = T_{\rm f} ~ ~ {3 \times 10^{-26} ~ {\rm cm}^3 ~ {\rm s}^{-1} \over \langle \sigma_{\rm ann} v \rangle} .
\ee
The final DM abundance will then be given:
\be
\label{annden}
{n_\chi \over s} = \left({n_\chi \over s}\right)_{\rm thr} ~ {3 \times 10^{-26} ~ {\rm cm}^3 ~ {\rm s}^{-1} \over \langle \sigma_{\rm ann} v \rangle} ~ \left(T_{\rm f} \over T_{\rm r} \right) .
\ee

This scenario can work well in the case of Higgsino DM, for which $\langle \sigma_{\rm ann} v \rangle > 3 \times 10^{-26} ~ {\rm cm}^3 ~ {\rm s}^{-1}$) as mentioned before, provided that the reheat temperature from $S$ decay satisfies Eq.~(\ref{anncond}).
}
\item{
{\bf Branching Scenario:} If Eq.~(\ref{anncond}) is not satisfied, then annihilation will be rendered ineffective. This happens if $T_{\rm r}$ is too low and/or $\langle \sigma_{\rm ann} v \rangle$ is too small.

The first possibility is that $\langle \sigma_{\rm ann} > 3 \times 10^{-26} ~ {\rm cm}^3 ~ {\rm s}^{-1}$, but $T_{\rm r}$ is lower than that given in Eq.~(\ref{anncond}). In this case non-thermal Higgsino DM must be produced via ``Branching Scenario''.

On the other hand, we note that Eq.~(\ref{anncond}) can never be satisfied if $\langle \sigma_{\rm ann} v \rangle < 3 \times 10^{-26} ~ {\rm cm}^3 ~ {\rm s}^{-1}$. It is seen from Eq.~(\ref{thr}) that this results in $(n_\chi/s)_{\rm thr} > (n_\chi/s)_{\rm obs}$ (hence ``thermal overproduction''). This leaves ``Branching Scenario'' as the only possibility for non-thermal DM production in this case. Bino DM provides a prime example of this case.

The final DM abundance will be the same as that produced from $S$ decay, which follows
\be \label{brden}
{n_\chi \over s} = {3 T_{\rm r} \over 4 m_S} ~ {\rm Br}_\chi .
\ee
}
\end{itemize}

\subsection{Challenges for non-thermal dark matter from modulus decay}

Modulus decay provides a natural scenario for non-thermal DM \cite{Moroi:1999zb}. Moduli heavier than 50 TeV decay before the onset of BBN, which allows their utilization as the source of DM production. The modulus decay rate is given by $\Gamma_S \sim m^3_S/2 \pi M^2_{\rm P}$ \cite{Allahverdi:2010rh}. For $m_S > 50$ TeV, this results in $Y_S \gsim 10^{-7}$.

``Annihilation Scenario'' can be realized in this framework. We have recently discussed non-thermal Higgsino DM from modulus decay in scenarios with mixed anomaly-modulus mediation of supersymmetry breaking \cite{Allahverdi:2012wb}. For a typical modulus mass $m_S \sim {\cal O}({\rm 1000})$ TeV in this scenario, one finds $T_{\rm r} \sim {\cal O}({\rm GeV})$. Eq.~(\ref{anncond}) can be satisfied for Higgsino annihilation cross sections that are compatible with Fermi bounds \cite{GeringerSameth:2011iw} with a mass $m_\chi \sim 100-1000$ GeV \cite{Allahverdi:2012wb}.

``Branching Scenario'', however, is not easily realizable. The fact that $Y_S \gsim 10^{-7}$ requires ${\rm Br}_\chi \leq 5 \times 10^{-5}$ for $m_\chi \geq 100$ GeV. Such a small ${\rm Br}_\chi$ may be obtained for two-body decays of the modulus \cite{Moroi:1999zb}, but three-body decays will inevitably set a lower bound ${\rm Br}_\chi \gsim 10^{-3}$. Obtaining the correct DM relic abundance within ``Branching Scenario'' then requires that $Y_S$ be further lowered, which may happen by means of geometric suppression \cite{Allahverdi:2010rh}.

An additional challenge is posed by gravitino production from modulus decay, whose abundance follows $(n_{3/2}/s) = Y_S ~ {\rm Br}_{3/2}$, where ${\rm Br}_{3/2}$ is the branching fraction for gravitino production from modulus decay. Gravitinos decay much later than the modulus. For example, in the scenario discussed in \cite{Allahverdi:2012wb} $m_{3/2} \sim {\cal O}(100)$ TeV, which implies they decay just before the onset of BBN. Gravitinos produce DM particles upon decay, which requires that
\be \label{gravdens}
{n_{3/2} \over s} < 5 \times 10^{-10} ~ \left({1 ~ {\rm GeV} \over m_\chi}\right).
\ee

For $m_\chi \geq 100$ GeV, this results in the bound $(n_{3/2}/s) < 5 \times 10^{-12}$. Since $Y_S \gsim 10^{-7}$, we then need ${\rm Br}_{3/2} < 10^{-5}$. However, in the simples example based on KKLT model \cite{Kachru:2003aw} we have ${\rm Br}_{3/2} \sim 10^{-2}$. One may lower ${\rm Br}_{3/2}$ by modifying the K\"ahler potential, but a successful implementation that does not affect other aspects of the non-thermal DM scenario is a non-trivial task.

To summarize, a completely successful non-thermal DM scenario from modulus decay is challenging, in particular in models with thermal overproduction (notably Bino DM). The reason being that ${\rm Br}_\chi$ and ${\rm Br}_{3/2}$ are typically too large in this case.

\section{Late Decay of a Visible Sector Scalar}

In this section, we present an explicit model of non-thermal DM from late decay of a visible sector field. We describe the model and its field content, and show how it can address the abovementioned issue with ${\rm Br}_{3/2}$.

\subsection{The Model}

The visible sector consists of the minimal supersymmetric standard model (MSSM) augmented with extra superfields:\\
\\
\noindent
$(i)$ A singlet $S$ whose decay is the origin of DM.\\
\\
\noindent
$(ii)$ Two flavors of iso-singlet color triplets $X_{1,2},~\bar{X}_{1,2}$ with hypercharges $+4/3,-4/3$ respectively.\\
\\
\noindent
$(iii)$ Two flavors of singlets $N_{1,2}$. \\

Multiple flavors of $X,~{\bar X},~N$ are introduced to accommodate late-time baryogenesis (which we will discuss in Section V).

We choose charge assignments under $R-$parity such that $R$-parity is conserved, hence LSP is stable. This implies that ${\tilde X},~{\tilde {\bar X}},~S$ and $N$ are $R$-parity even, while their SUSY partners are $R$-parity odd. We also assume that the lightest MSSM neutralino is the DM candidate.

The superpotential of the visible sector is $W_{\rm visible} = W_{\rm MSSM} + W_{N,X} + W_{S}$, where
\bea\label{superpot1}
W_{N,X} =  \lambda^\prime d^c d^c {\bar X} + \lambda N u^c X + {m_{N} \over 2} N N +  m_{X} X \overline{X} \, , \nonumber \\
 \,
\eea
and
\be\label{superpot2}
W_{S} = h S  X {\bar X} + \frac{1}{2} M_S S^2 \,\, .
\ee
For simplicity, we have omitted the flavor and color indices. Henceforth, we use the same symbol for superfields and their corresponding $R$-parity even component fields, while $R$-parity odd fields are distinguished by a $\tilde {}$.

We note that some terms that are gauge-invariant are absent in $W_S$, namely $S N u^c,~N^3,~S N N,~S^3$. The first two terms are forbidden by $R$-parity, while invoking some other discrete or continuous symmetry (like an $R$-symmetry) may help forbid or suppress the last two terms.

Soft SUSY breaking terms typically make scalar components of chiral superfields heavier than their fermionic counterparts. In the case of $S$ superfield, one has $m^2_S = M^2_S + m^2_{\rm soft} \pm B M_S$, where $m_{\rm soft}$ denotes the soft mass of $S$ and $B$ is the $B$-term associated with the superpotential mass term for superfield $S$. We assume the following mass condition
\bea \label{massconditions}
m_{N} & < & m_S \ll m_{X} \, , \nonumber \\
m_S & > & 2 m_\chi \, .
\eea
It is also reasonable to assume that all $R$-parity odd colored particles have a mass larger than $m_S$. The reason being that soft breaking masses of these particles are driven toward large values at low energies by radiative corrections with $SU(3)_C$ gauge interactions while $S$ is a singlet.

These conditions imply that $S$ cannot decay to either of $X,~{\tilde X}$ or ${\bar X},~{\tilde {\bar X}}$ fields. Moreover, it cannot decay to any $R$-parity odd colored fields. Similarly, decay to $R$-parity odd scalars is kinematically blocked since these masses are governed by the soft terms, which supposedly have the smallest value for a singlet field. $S$ can however decay to DM particles, which is an essential part of this model.

\begin{figure}[ht]
\centering
\includegraphics[width=2.0in]{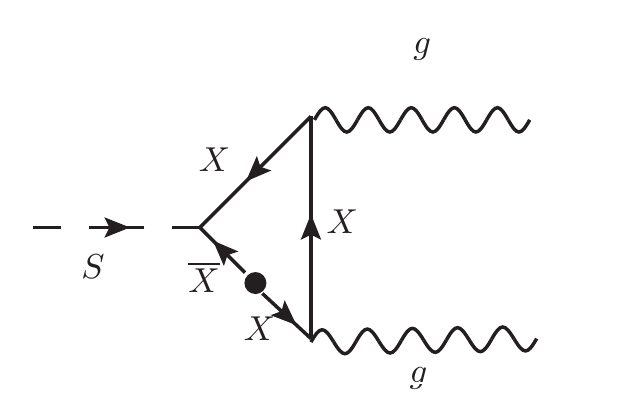}
\caption{Diagram showing the dominant one-loop decay of $S$. There is a similar diagram with ${\tilde X},~{\tilde {\bar X}}$ running in the loop.}
\label{visiblecladoSdecaytogluons}
\end{figure}

As a consequence, $S$ dominantly decays into two gluons through the one-loop diagram shown in Figure \ref{visiblecladoSdecaytogluons}.\footnote{$S$ can also decay into four-body final states containing $(N,~u^c,~
d^c)$ through mediation of off-shell ${\tilde X},~{\tilde {\bar X}}$. However, these decays are suppressed compared to the two-body decays by phase space factors and additional powers of $m_S/m_X$, and hence can be neglected. There are also subdominant decay modes of $S$ that proceed via loop diagrams and are important for DM production and baryogenesis. We will discuss these modes later.} The corresponding decay is given by
\be \label{Stogluons}
\Gamma_{S \rightarrow gg} \, \sim \, 2 \times \frac{0.17}{8\pi}\left(\frac{h g_3^2}{4\pi^2}\right)^2 \left(\frac{m_S}{m_X}\right)^2 m_S  \,\,\,.
\ee
Here $g_3 \sim 1$ is the $SU(3)_C$ gauge coupling constant and the two flavors of $X$ are taken into account. The precise expression for $\Gamma_{S \rightarrow gg}$ is given in the Appendix.

The fact that $S$ decay is loop suppressed combined with $m_S \ll m_X$ help us obtain a sufficiently low reheat temperature $T_{\rm r} \sim (\Gamma_S M_{\rm P})^{1/2} \ll T_{\rm f}$. For example, for $m_S \sim 1$ TeV and $m_X \sim 10-100$ TeV, we find $T_{\rm r} \sim {\cal O}({\rm GeV})$ if $h \sim 10^{-7}-10^{-6}$.

The magnitude of $h$ needed is similar to the electron Yukawa coupling, and also comparable to typical values of neutrino Dirac Yukawa couplings in TeV scale see-saw models. Its smallness may be explained in different ways. There may exist some symmetry, broken at the scale of grand unified theories (GUT), under which $S$ is charged. Then the term $S X {\bar X}$ arises from a higher-order $M_{\rm P}$ suppressed operator after symmetry breaking, and its strength will be suppressed by powers of $M_{\rm GUT}/M_{\rm P}$. Moreover, see Eq.~(\ref{Stogluons}), we notice that the combination $h/m_X$ appears in $\Gamma_S$, which acts like an effective coupling. Therefore one can make $h$ larger by simultaneously increasing $m_X$ with the same factor. For example, we find $h \sim {\cal O}(1)$ if $m_X \sim 10^{11}$ GeV.

We also note that for a gravitationally interacting field $S$ the decay rate will be $\propto m^3_S/M^2_{\rm P}$. It is seen from Eq.~(\ref{Stogluons}) that $S$ decay to gluons occurs with a strength much larger than that for a gravitational decay as long as $4 \pi^2 m_X/h \ll M_{\rm P}$. This is clearly the case for $m_S \sim {\cal O}({\rm TeV})$ and $T_{\rm r} > 3$ MeV, which justifies $S$ belonging to the visible sector rather than a hidden sector with gravitationally suppressed coupling to matter.

\subsection{Gravitino Production}

Gravitinos can be produced from $S$ decay if $m_{3/2} < m_S$. Considering that $m_S \sim 1$ TeV, this is kinematically possible only if $m_{3/2} \ls {\cal O}({\rm TeV}$. This implies that there will be {\it no} gravitino production form $S$ decay for a very wide mass range $m_{3/2} > {\cal O}({\rm TeV})$.

When kinematically possible, the important decay mode is $S \rightarrow {\tilde G} + {\tilde S}$, whose decay width is given by \cite{Moroi:1995fs}
\be \label{Stograv}
\Gamma_{S \rightarrow {\tilde G} + {\tilde S}} \sim {1 \over 48 \pi} ~ {m^3_S \over M^2_{\rm P}} .
\ee

Gravitinos with ${\cal O}({\rm TeV})$ mass decay long after BBN. Their abundance should be low enough in order not to ruin successful predictions of BBN. Since $m_{3/2} < m_S$, and $S$ is assumed to be lighter than all colored superparticles, gravitino decay modes can only be radiative. Successful BBN in this case requires that \cite{BBN2}
\be \label{bbnbound}
{n_{3/2} \over s} < 10^{-12} .
\ee
This also insures that gravitino decay will not overproduce DM particles with a mass $m_\chi \lsim 500$ GeV.

The abundance of gravitinos produced from $S$ decay follows
\be \label{reheatbound}
{n_{3/2} \over s} = Y_S ~ {\Gamma_{S \rightarrow {\tilde G} + {\tilde S}} \over \Gamma_{S \rightarrow g g}} \sim {1 \over 48 \pi} ~ \left({m_S \over M_{\rm P}}\right)^2 ~ \left({M_{\rm P} \over T_{\rm r}}\right) .
\ee
For $m_S \sim {\cal O}({\rm TeV})$ the limit in Eq.~(\ref{bbnbound}) is satisfied provided that $T_{\rm r} \gsim 2.5$ MeV. We see that gravitino production will not be a problem for the entire range of $T_{\rm r}$ allowed by BBN (i.e., $T_{\rm r} \gsim 3$ MeV). Moreover, this condition will be irrelevant altogether if $m_{3/2} > m_S$, in which case $S$ decay to the gravitino will be forbidden kinematically.

The fact that that gravitino overproduction from $S$ decay is avoided comes as a direct consequence of $S$ mainly decaying into visible sector fields (i.e., gluons), while its decay to gravitinos is $M_{\rm P}$ suppressed. The situation is very different when $S$ is a modulus field, as pointed out earlier, since all decay modes are gravitationally suppressed in that case.

\section{Dark Matter from Visible Sector Decay}

In this section we show how the model presented in the previous section can produce the observed DM relic abundance via the ``Branching Scenario'' and ``Annihilation scenario'' both. We also discuss successful non-thermal production of Higgsino and Bino DM from visible sector decay. Recently, Higgsino type LSP has attracted significant attention from the natural SUSY perspective \cite{Papucci:2011wy, Hall:2011aa, Baer:2012uy, Allahverdi:2012wb, Gogoladze:2012yf, Baer:2012cf}. However, if the Higgsino mass is in the sub-TeV region, the annihilation rate is larger than the nominal value $3 \times 10^{-26}$ cm$^3$ s$^{-1}$, which yields insufficient thermal relic abundance. The concern with this scenario is that the current constraint from the Fermi-LAT data does not allow much room for the annihilation cross-section to be too large. These results appear to prefer smaller annihilation cross-sections for smaller values of LSP mass. This prefers Bino type LSP, which will have a thermally overproduced abundance.

\subsection{Branching Fraction for Decay to Dark Matter Particles}

The decaying scalar $S$ has $R$-parity charge +1, which implies that it can only decay to an even number of $R$-parity odd particles. If it was $R$-parity odd instead, all of the decay modes would produce $R$-parity odd particles, thus resulting in ${\rm Br}_\chi = 1$. However, one can now obtain a small ${\rm Br}_\chi$ if $S$ decay to $R$-parity odd particles is suppressed relative to that of $S$ decay to gluons.

\begin{figure}[ht]
\centering
\includegraphics[width=2.5in]{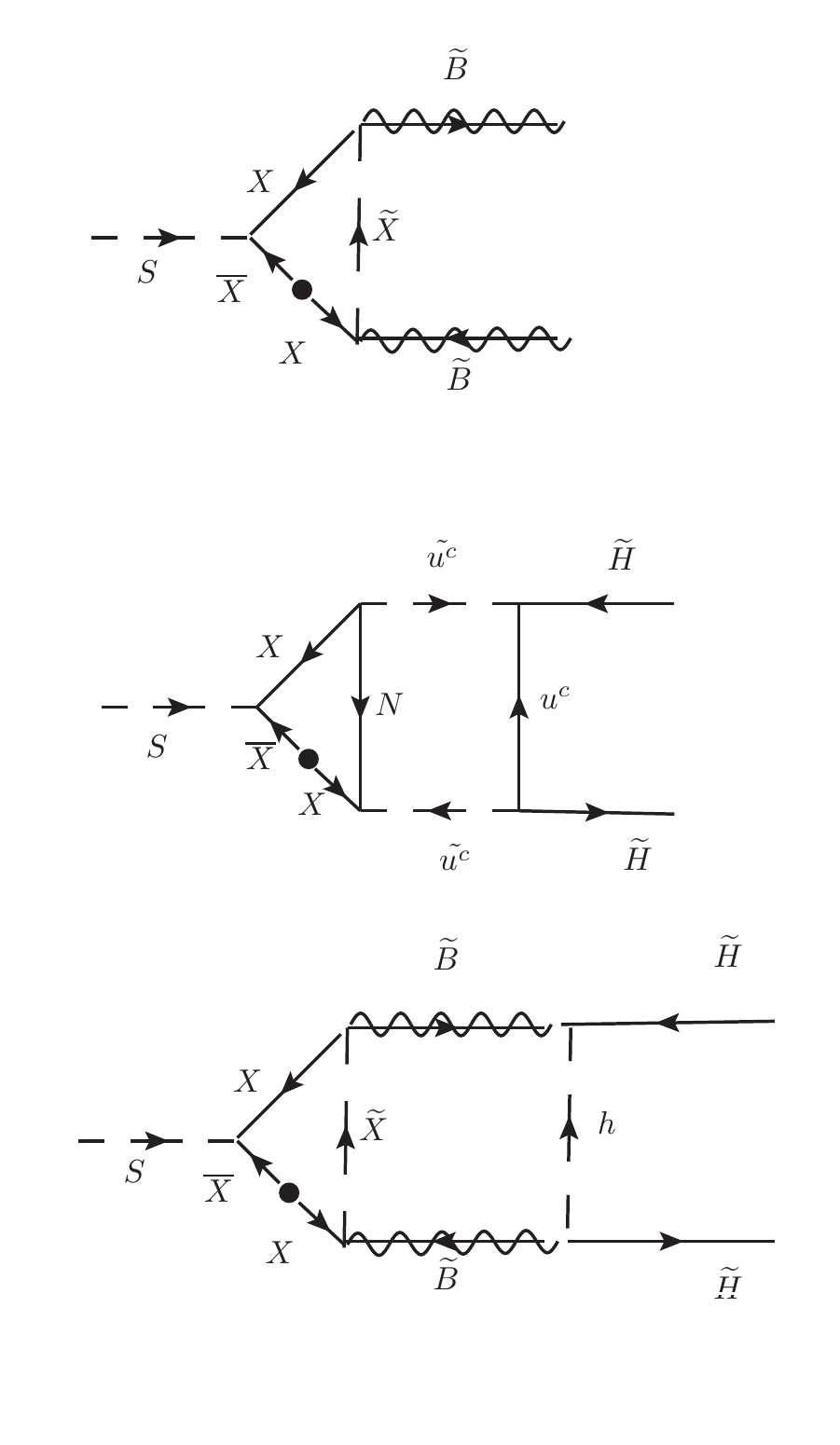}
\caption{Dominant diagrams for the decay of $S$ to Bino (top) and Higgsino (middle and bottom). There are additional diagrams that are obtained by switching internal lines to their SUSY partners.}
\label{visiblecladoStoDarkMatter}
\end{figure}

The dominant mode for producing $R$-parity odd particles is $S \rightarrow {\tilde B} {\tilde B}$ that proceeds through the one-loop diagram shown in Figure~\ref{visiblecladoStoDarkMatter} (top diagram). This mode is kinematically allowed if Bino is the DM particle, see Eq.~(\ref{massconditions}). The corresponding decay width is given by (for more details see the Appendix)
\bea \label{Stobino}
\Gamma_{S \rightarrow \widetilde{B}\widetilde{B}} \, & \sim & \, 2 \times \frac{0.12}{8\pi} ~ \left(\frac{h Y^2_X g_1^2}{16\pi^2}\right)^2  \left(\frac{m_{\widetilde{B}}}{m_X}\right)^2 m_S \, ,
\eea
where $g_1$ is the $U(1)_Y$ gauge coupling constant. Both flavors of $X$ have been taken. After using Eq.~(\ref{Stogluons}), this results in the following branching fraction
\be \label{binobranch}
{\rm Br}_{\chi} \sim 0.71 ~ \left({Y_X g_1 \over 2 g_3}\right)^4 ~ \left({m_{\chi} \over m_S}\right)^2  ~ ~ ~ ~ ~ ({\rm Bino ~ LSP}) .
\ee
Inserting numerical values for the parameters $m_S \sim 1$ TeV, $m_{\tilde B} \sim 100$ GeV, $g_1 \sim 0.3$, $g_3 \sim 1$, and $Y_X = 4/3$, we find ${\rm Br}_{\chi} \sim 1.1 \times 10^{-5}$. This is much smaller than what one finds from modulus decay ${\rm Br}_\chi \gsim 10^{-3}$ \cite{Allahverdi:2010rh}.

If the LSP is not purely Bino but rather mixed Higgsino/Bino or Wino/Bino, the branching fraction will be
\be \label{mixedbranch}
{\rm Br}_{\chi} \sim 0.71 ~ f^4 ~ \left({Y_X g_1 \over 2 g_3}\right)^4 ~ \left({m_{\chi} \over m_S}\right)^2  ~ ~ ~ ~ ~ ({\rm Mixed ~ LSP}) ,
\ee
where $f$ is the Bino fraction of the DM.

For a Higgsino LSP the branching fraction in Eq.~(\ref{mixedbranch}) will be vanishingly small since $f \ll 1$. However, Higgsinos can be directly produced from $S$ decay at two-loops through diagrams shown in Figure~\ref{visiblecladoStoDarkMatter} (middle and bottom diagrams). The decay widths for the two-loop diagrams are
\bea \label{Stohiggsino}
\Gamma_{S \rightarrow {\tilde H} {\tilde H}} & \sim & \frac{1}{8\pi} ~ \left({h \lambda^2_t \over 4 \pi^2 }\right)^2 ~ \left({y^2_t \over 4 \pi^2}\right)^2 ~ \left({m_{\tilde H} \over m_X}\right)^2 m_S \, , \nonumber \\
\Gamma_{S \rightarrow {\tilde H} {\tilde H}} & \sim & \frac{1}{8\pi} ~ \left({h Y^2_X g^2_1 \over 16 \pi^2 }\right)^2 ~ \left({Y^2_H g^2_1 \over 16 \pi^2}\right)^2 ~ \left({m_{\tilde H} \over m_X}\right)^2 m_S \, , \nonumber \\
& & \,
\eea
where $y_t \sim 1$ is the top Yukawa coupling, $\lambda_t$ is the coefficient of $N t^c X$ term in Eq.~(\ref{superpot1}), and $Y_H = 1$ is the Higgs hypercharge.

Unless $\lambda_t$ is very small, the first expression in Eq.~(\ref{Stohiggsino}) will be dominant. Then the branching fraction for Higgsino DM will be
\be \label{higgsinobranch}
{\rm Br}_\chi \sim {1 \over 16 \pi^4} ~ \left({\lambda_t \over g_3}\right)^{4} ~ \left({m_\chi \over m_S}\right)^2 ~ ~ ~ ~ ~ ({\rm Higgsino ~ LSP}) .
\ee
Much smaller values of ${\rm Br}_\chi$ can be obtained in the case of Higgsino DM. For example, we can find ${\rm Br}_\chi \sim 10^{-10}-10^{-4}$ for $m_S \sim 1$ TeV, $m_{\tilde H} \sim 100-500$ GeV, and $\lambda_t \sim 0.1-1$.

We note that the second expression in Eq.~(\ref{Stohiggsino}) takes over for $\lambda_t \ll 0.1$, which provides a lower bound on ${\rm Br}_\chi$ regardless of how small $\lambda_t$ is.

\subsection{Higgsino Dark Matter}

In Table~\ref{benchparameters}, we show the model parameters and typical mass scales considered in this work. In Table~\ref{summarytable}, we give a summary of the various possibilities for Higgsino and Bino DM. 

For a purely Higgsino DM, ${\rm Br}_\chi$ is given by Eq.~(\ref{higgsinobranch}). Since Higgsino LSP is thermally underproduced, the correct relic abundance can be obtained via both of the ``Annihilation'' and ``Branching'' scenarios discussed in Section II. The important factor in determining which of the scenarios can work is the size of coupling $\lambda_t$.

To illustrate this, we first consider the case with $\lambda_t \sim 1$. Then Eq.~(\ref{higgsinobranch}) results in ${\rm Br}_\chi \gsim 10^{-5}$ for $m_{\tilde H} \geq 100$ GeV. Combined with the condition for not overproducing gravitinos from $S$ decay, see Eq.~(\ref{reheatbound}), this leads to $Y_S ~ {\rm Br}_\chi \gsim 10^{-10}$. Therefore the observed relic density cannot be obtained in the ``Branching Scenario''. This leaves the ``Annihilation Scenario'' as the only possibility when $\lambda_t$ is large. This is shown in Table~\ref{summarytable}. For $\lambda_t \sim 1$, ``Annihilation Scenario" is the only option even for $T_r$ at the BBN bound. The correct relic density may be obtained in ``Annihilation Scenario" for $T_r \sim 0.5$ GeV \cite{Allahverdi:2012wb}.

Next, we consider the case with $\lambda_t \sim 0.2$. Then we see from Eq.~(\ref{higgsinobranch}) that ${\rm Br}_\chi \sim 10^{-8}-10^{-7}$ for $m_{\tilde H} \sim 100-500$ GeV. This implies that the correct relic abundance via the ``Branching Scenario'' can be obtained for $10 ~ {\rm MeV} < T_{\rm r} < 1 ~ {\rm GeV}$. A typical example of this scenario is shown in Table~\ref{summarytable}.

The flexibility that one can make either of the ``Annihilation'' and ``Branching'' scenarios work by dialing $\lambda_t$ is also important from another point of view. ``Annihilation Scenario'' relies on a large annihilation rate $\langle \sigma_{\rm ann} v \rangle$. Therefore, tightening of the Fermi bounds on $\langle \sigma_{\rm ann} v \rangle$ \cite{Hooper:2012sr}, will put increasing pressure on this scenario. It is therefore important that one can have a viable ``Branching Scenario'', regardless of the improving constraints, by lowering $\lambda_t$.

\begin{table}[!htp] 
\caption{Typical scales and values of model parameters in the scenarios considered in this work. The reheat temperature is $T_r \sim \mathcal{O}(1)$ GeV for the values presented in the Table.}
\label{benchparameters}
\begin{center}
\begin{tabular}{c c} \hline \hline
  Parameter      & Value   \\ \hline \hline \\
  $m_S$       & $\mathcal{O}({\rm TeV})$   \\
  $m_{X_1} \sim m_{X_2}$       & $\mathcal{O}(10 \,\, {\rm TeV})$   \\
  $m_{N_1}$       & sub-TeV   \\
  $\lambda^\prime$       & $\mathcal{O}(1)$   \\
  $\lambda_{t}$ & $\mathcal{O}(0.1-1)$  \\
  $h$ & $\mathcal{O}(10^{-6})$  \\  \hline \hline
\end{tabular}
\end{center}
\end{table}
\begin{table}[!htp] 
\caption{Summary of the various scenarios considered in this work, for typical values of model parameters. We have taken $m_S = 1$ TeV.  For Higgsino DM, both ``Annihilation Scenario"  and ``Branching Scenario" may be obtained depending on the value of $\lambda_t$. For Bino DM, successful ``Branching Scenario" requires light DM and low reheat, which may be obtained by suitably choosing $h/m_X$. All mass scales are shown in GeV.}
\label{summarytable}
\begin{center}
\begin{tabular}{c c c c c} \hline \hline
  DM      & Mass        & $T_r$       & ${\rm Br}_\chi$  & Scenario  \\ \hline \hline \\

          & $100$   \,\,    & $0.5$    \,\,      & $6 \cdot 10^{-6}$  & Annihilation ($\lambda_t = 1$) \\
         $\widetilde{H}$  &              &      &  & \\
          & $100$   \,\,    & $0.5$     \,\,     & $1 \cdot 10^{-8}$  & Branching ($\lambda_t = 0.2$)\\ 
\hline \\

$\widetilde{B}$  & $60$ \,\,  & $3 \cdot 10^{-3}$ \,\, &  $4 \cdot 10^{-6}$ \,\,\,& Branching ($\frac{h}{m_X} \sim 10^{-14}$)
\\
 \\  \hline \hline
\end{tabular}
\end{center}
\end{table}

\subsection{Bino Dark Matter}

Bino DM is thermally overproduced. This, as pointed before, leaves ``Branching Scenario'' as the only possibility for producing the correct DM density. In this case $(n_\chi/s) = Y_S ~ {\rm Br}_\chi$, where $Y_S \gsim 2 \times 10^{-6}$ for $m_S \sim 1$ TeV. Obtaining the observed relic density then requires that ${\rm Br}_\chi \leq 2.5 \times 10^{-4}/m_{\tilde B}$. We see from Eq.~(\ref{binobranch}) that this can be found for a Bino on the lighter side $m_{\tilde B} \sim 60$ GeV when the reheat temperature is close to its lower value from BBN bound $T_{\rm r} \sim 3$ MeV.

Having a viable scenario for non-thermal Bino DM is important. With the Fermi bounds on the annihilation rate improving \cite{GeringerSameth:2011iw}, Higgsino DM may be ruled out specially at the lower end of the sub-TeV mass range. This will motivate Bino DM, for which one can find a successful non-thermal scenario based on $S$ decay.

\section{Baryogenesis}

The entropy released in $S$ decay dilutes any previously produced baryon asymmetry, see Eq.~(\ref{entropy}), thus necessitating mechanisms of baryogenesis at temperatures far below the electroweak symmetry breaking scale. This can be achieved by introducing baryon number and $CP$ violating operators in a suitable extension of the SM~\cite{Allahverdi:2010im,Babu:2006wz}. Late decay of $S$ naturally provides the necessary departure from thermal equilibrium.

\begin{figure}[ht]
\centering
\includegraphics[width=3.5in]{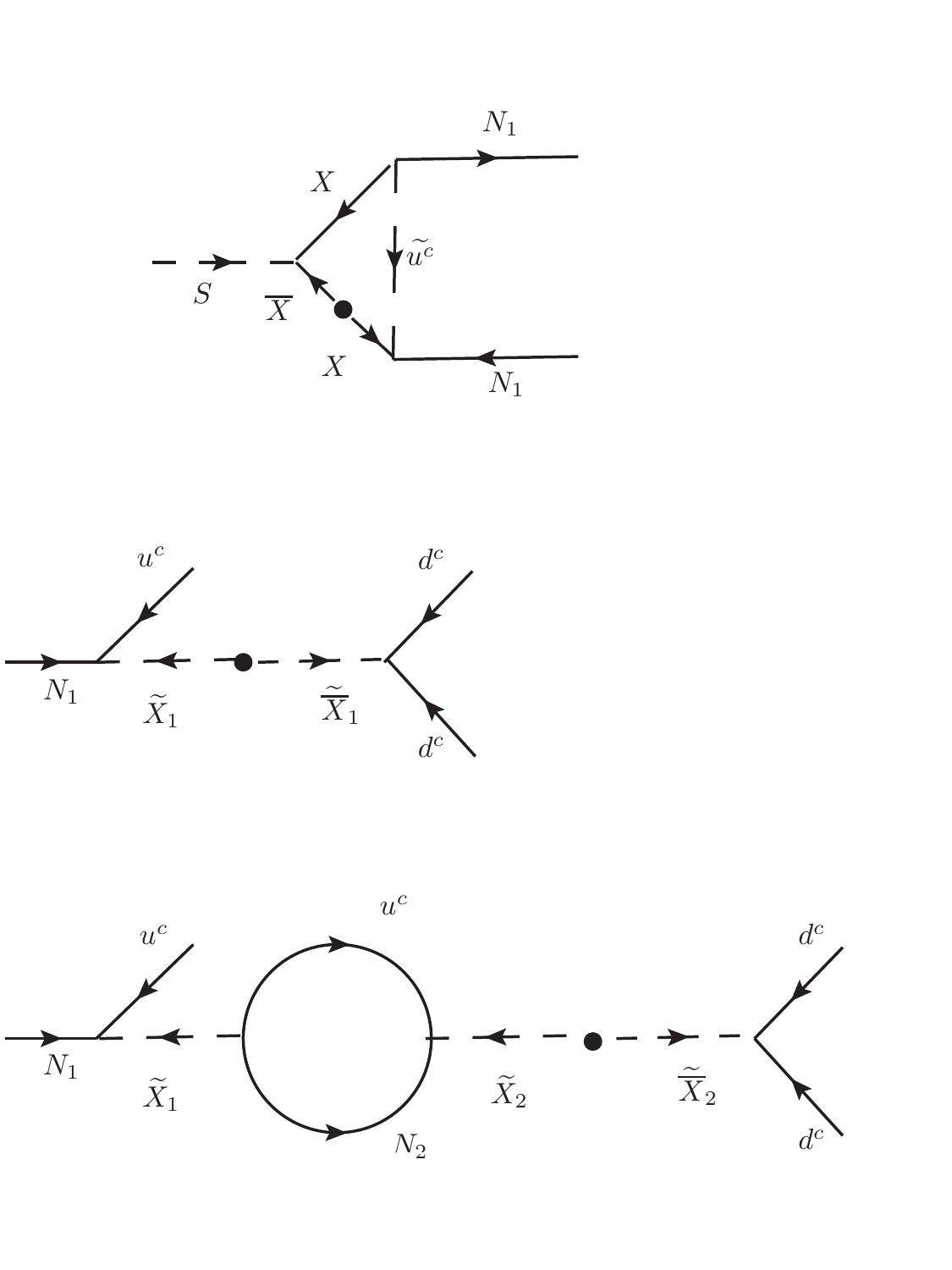}
\caption{Production of $N_1$ from $S$ happens at the one-loop level (top diagram). $N_1$ subsequently decays and generates baryon asymmetry (middle and bottom diagrams).}
\label{visiblecladoSdecayBaryogenesis}
\end{figure}

The model given in Eq.~(\ref{superpot1}) can give rise to low-temperature baryogenesis via $N$ decay~\cite{Allahverdi:2010im}. The decay of $S$ produces $N$ quanta through the one-loop diagram shown in Fig.~\ref{visiblecladoSdecayBaryogenesis} (top diagram). The rate for $S$ decay to the heavier of the two $N$ flavors, denoted by $N_1$, is
\be \label{StoN1N1}
\Gamma_{S \rightarrow N_1 N_1} \, \sim  \frac{1}{8\pi} ~ \left(\frac{h \lambda^2_u}{4\pi^2}\right)^2 ~ \left(\frac{m_{N_1}}{m_X}\right)^2 ~ m_S \,\,
\ee
where $\lambda_u$ is the coefficient of $N u^c X$ term. We have chosen $\lambda_u > \lambda_{c,t}$, so that the loop containing $u^c$ dominates the process. We note that $S$ decay to $N_2$ and quarks is subdominant due to the helicity suppression of fermionic final states, while decay to scalar ${\tilde N}$ and squarks is kinematically blocked, see Eq.~(\ref{massconditions}) and the subsequent discussion.

$N_1$ decay via the diagram in Fig.~\ref{visiblecladoSdecayBaryogenesis} generates baryon asymmetry, whose density is given by
\be \label{baryonovers}
\eta_{\rm B} \, \equiv {n_{B} - n_{\bar B} \over s} = Y_S ~ {\rm Br}_{N} ~ \epsilon \,\, ,
\ee
where ${\rm Br}_N$ is the branching fraction for producing $N_1$ from $S$ decay, and $\epsilon$ is the asymmetry parameter in $N_1$ decay. ${\rm Br}_N$ is given by
\be \label{StoNbbranch}
{\rm Br}_{N_1} \, \sim \, \left(\frac{\lambda_u}{g_3}\right)^4 ~ \left(\frac{m_{N_1}}{m_S}\right)^2  \,\,.
\ee
For $\lambda_u \sim 0.5$, we find ${\rm Br}_{N_1} \, \sim \, 0.01 $.

Taking $m_{X_1} \approx m_{X_2}$, similar masses for $X_{1,2}$ fermions and scalars, $\lambda^{\prime} \, \sim \, \mathcal{O}(1)$ for all flavors and colors (in accordance with Table~I), and $CP$ violating phases of $\mathcal{O}(1)$, the asymmetry parameter is given by
\be \label{epsilon}
\epsilon \, \sim \, \frac{1}{8\pi} ~ \left(\frac{m_{N_1}}{m_{X}}\right)^2 ~ \lambda^2_u \times \mathcal{O}(10)\,\,\, .
\ee
The $\mathcal{O}(10)$ factor comes from summing over three flavors of quarks plus including diagrams where $X_1$ are switched $X_2$. Then, for $\lambda_u \sim 0.5$, we get $\epsilon \, \sim \, 10^{-5} \,\,$.

It is seen from Eqs.~(\ref{baryonovers},~\ref{StoNbbranch},~\ref{epsilon}) that we can obtain the desired value $\eta_{\rm B} \sim 10^{-10}$ for $\lambda_u \sim 0.5-1$, $M_{N_1} \sim 500$ GeV, $m_S \sim {\cal O}({\rm TeV})$, $T_{\rm r} \sim 10 ~ {\rm MeV}-1$ GeV, and $m_X \sim 10-100$ TeV.

One comment is in order at this point. $N_2$ fermions produced in $N_1$ decay themselves to three quarks. This must happen well before the onset of BBN in order not to ruin its success. The decay rate of $N_2$ is given by \cite{Babu:2006wz}
\be
\Gamma_{N_2} \, \sim \,\frac{C}{126} ~ \frac{1}{192\pi^3} ~ \left(\frac{B m^2_{N_2}}{m^3_X}\right)^2 ~ m_{N_2} \,\,\, ,
\ee
where $B$ is the $B-$term associated with $m_X X {\bar X}$ mass term in Eq.~(\ref{superpot1}), and $C \sim 6$ is a color multiplicity factor. One can see that $N_2$ decay well before 0.1 s for $m_{N_2} \gsim 100$ GeV, $B \sim {\cal O}({\rm TeV})$, and the parameter values given in Table I.

Finally, we note that the ``Branching Scenario'' for non-thermal DM production (like in the case of Bino LSP) provides a visible sector realization of ``Cladogenesis'' mechanism proposed in~\cite{Allahverdi:2010rh} to address the baryon-DM coincidence puzzle. It is clear from our discussions that this scenario is a considerably more flexible than that when modulus decay is the source of non-thermal DM~\cite{Allahverdi:2010rh,Allahverdi:2012wb}.

\section{Conclusion}

In large regions of SUSY parameter space, the DM annihilation cross-section is larger (as for Higgsino or Wino LSP) or smaller (as for Bino LSP) than the nominal value of $3\times 10^{-26}$cm$^3$ s$^{-1}$ in the thermal WIMP scenario. Non-thermal scenarios of DM production will be needed in these cases to yield the correct relic abundance, while being compatible with the bounds from indirect DM searches on the DM annihilation cross section particularly the Fermi-LAT results.

Motivated by these considerations, in this paper, we have presented a non-thermal scenario that relies on the visible sector of a SUSY model. In this model the late decay of an $R$-parity even scalar field $S$ that is a SM singlet (but may be charged under a higher rank gauge group) produces DM. $S$ is coupled to new colored fields $X,~{\bar X}$ with a mass relation $m_S \ll m_X$. Assuming that all $R$-parity odd colored fields are heavier than $S$, it will dominantly decay into gluons at the one-loop level. The combination of a small coupling $h$ between $S$ and $X,~{\bar X}$, the mass relation $m_X \gg m_S$, and the one-loop factor can naturally lead to a late decay of $S$ that yields a low reheat temperature $T_{\rm r}$. As we saw above, one can obtain $T_{\rm r} \sim {\cal O}({\rm GeV})$ for $m_S \sim 1$ TeV, $m_X \sim 50$ TeV, and $h \sim 10^{-6}$.

The branching fraction for $S$ decay to DM particles ${\rm Br}_\chi$ depends on the nature of the LSP. If the LSP is Bino, then $S$ can decay to a pair of DM particles at the one-loop level, where ${\rm Br}_\chi \gsim 10^{-6}$ for $m_\chi \geq 60$ GeV. For a Higgsinos LSP, the same decay occurs at the two-loop level yielding ${\rm Br}_\chi \sim 10^{-10}-10^{-5}$ (the exact value depending on the model parameters) for $m_\chi \sim 100-500$ GeV. As a consequence, one can obtain the observed relic abundance in both larger and smaller annihilation cross-section regions of SUSY parameter space.

For Higgsino DM, the correct relic density can be found directly from $S$ decay if ${\rm Br}_\chi$ is sufficiently small. The DM annihilation cross-section will be irrelevant in this case (hence ``Branching Scenario''). We showed a benchmark point for this scenario where $m_\chi \sim 100$ GeV and $T_{\rm r} \sim {\cal O}({\rm GeV})$. For larger values of ${\rm Br}_\chi$, Higgsinos will undergo residual annihilation upon production from $S$ decay (hence ``Annihilation Scenario''). This scenario can yield the correct relic abundance if the annihilation rate is larger than $3 \times 10^{-26} ~ {\rm cm}^3 ~ {\rm s}^{-1}$ by a factor of $(T_{\rm f}/T_{\rm r})$.

In the case of Bino DM, where the annihilation cross-section is small, ``Branching Scenario'' is the only option. The correct relic density is obtained for a lighter Bino $m_\chi \lsim 100$ GeV and smaller values of reheat temperature $T_{\rm r} \sim 3$ MeV. One can reduce $T_{\rm r}$ by lowering the coupling $h$ and/or raising $m_X$. A viable scenario for non-thermal Bino DM is particularly important in light the Fermi-LAT result, which appear to prefer smaller annihilation cross-sections for smaller values of DM mass.

The above scenarios are summarized in Table~\ref{summarytable}.

The scenario presented in this paper has two main advantages over that using moduli decay. First, gravitino production from $S$ decay is naturally suppressed by the virtue of $S$ belonging to the visible sector. Second, ${\rm Br}_\chi$ can be made sufficiently small by dialing the model parameters, which is essential for a successful realization of the ``Branching Scenario'' (the only possibility in the case of Bino DM). To attain these virtues in scenarios of non-thermal DM from moduli decay, one needs to satisfy non-trivial conditions on the K\"ahler geometry of the underlying effective supergravity theory.

Finally, our model can also successfully generate the baryon asymmetry of the universe. $S$ decay also produces the SM singlets $N$ at the one-loop level, whose subsequent decay via baryon number and $CP$ violating couplings to the heavy colored states $X,~{\bar X}$ can yield the desired asymmetry $\eta_{\rm B} \sim 10^{-10}$. Moreover, the baryon-DM coincidence puzzle can be addressed in the context of the ``Branching Scenario'', since both baryon asymmetry and DM abundance are directly produced from $S$ decay in this case.

\section{Acknowledgement}

This work is supported in part by the DOE grant DE-FG02-95ER40917.

\appendix

\section{Calculations of $S$ Decay Width}

We first discuss the decay of $S$ into two gluons. There is a contribution from the loop of fermions shown in Fig.~\ref{visiblecladoSdecaytogluons}, as well as a loop of sfermions. The total decay width is given by \cite{Djouadi:2001kba}
\bea
\Gamma_{S \rightarrow gg} &=& \frac{1}{9\sqrt{2}}\frac{1}{8\pi}\left(\frac{h g_3^2}{4\pi^2}\right)^2 \left(\frac{m_S}{m_X}\right)^2 m_S \nonumber \\
&\times & \left(\frac{3}{4}\mathcal{A}_{1/2}(\tau) + \frac{3}{4} \mathcal{A}_{\rm SUSY}(\tau)\right)^2 \,\,,
\eea
where
\bea
\mathcal{A}_{1/2}(\tau) &=& 2\left(\tau + (\tau -1)f(\tau)\right)\tau^{-2} \nonumber \\
\mathcal{A}_{\rm SUSY}(\tau) &=& -\left(\tau - f(\tau)\right)\tau^{-2} \nonumber \\
f(\tau) &=&   \arcsin ^2 \sqrt{\tau}
\eea
and
\be
\tau = \frac{m^2_S}{4m^2_{\rm loop}} \,\,.
\ee
Here, $m_{\rm loop}$ is the mass of the particle in the loop. The amplitudes $\mathcal{A}_{1/2}$ and $\mathcal{A}_{\rm SUSY}$ refer to the fermion and sfermion loops respectively.

With the assumed mass conditions, one obtains the decay width given in Eq.~(\ref{Stogluons}), which further accounts for the two flavors of $X$.

The decay width into Binos is given by
\be
\Gamma_{S \rightarrow \widetilde{B}\widetilde{B}} = \left(\frac{1}{4\pi^2}\right)^2 N^2_C \delta^2_S \,\, ,
\ee
where
\bea
\delta_S &=&  \frac{h}{2\sqrt{2}} 4 \pi \alpha \left(\frac{Y_X}{2}\right)^2 \,  m_X m_{\widetilde{B}} \times \nonumber \\
&& \left[C_0(m_X) - 2C^+_1(m_X) \, + \, C^+_1(m_{\widetilde{X}}) \right] \nonumber
\eea
In the above equation, $C_0$ and $C^+_1$ are Veltman-Passarino functions \cite{Passarino:1978jh}. With the assumed mass conditions, this expression is reduces to  Eq.~(\ref{Stobino}).

\end{document}